\documentclass[aps,nofootinbib,floatfix,noshowpacs,preprint,10pt,]{revtex4}
\usepackage{mathrsfs}
\usepackage{graphicx}
\usepackage{epstopdf}
\usepackage{amsmath}
\usepackage{amsfonts}
\usepackage{amssymb}
\usepackage{color}
\usepackage{siunitx}

\usepackage{booktabs}
\usepackage{array}
\usepackage{epsfig}
\usepackage{CJK}
\usepackage{eepic}
\usepackage{bbm}
\usepackage{dcolumn}
\usepackage{bm}
\usepackage{ulem}
\usepackage{slashbox}
\usepackage{multirow}
\usepackage{CJKfntef}
\usepackage{caption}
\usepackage{subfigure}

\newcommand{\omits}[1]{}

\def\bc{\begin{center}}

\def\ec{\end{center}}
\def\be{\begin{eqnarray}}
\def\ee{\end{eqnarray}}

\definecolor{dyellow}{rgb}{1.,0.8,.0}
\definecolor{myblue}{rgb}{.1,.1,.7}
\definecolor{dcyan}{rgb}{.0,.6,.6}
\definecolor{cyan}{rgb}{0.4,1.0,1.0}
\definecolor{dmagenta}{rgb}{0.6,0.0,0.6}
\definecolor{brown}{rgb}{0.6,0.2,0.}
\definecolor{darkblue}{rgb}{.0,.0,0.5}
\definecolor{darkred}{rgb}{0.75,0.0,0.0}
\definecolor{orange}{rgb}{1.,.6,.0}
\definecolor{dorange}{rgb}{0.8,.4,.0}
\definecolor{green}{rgb}{0.0,1.0,0.0}
\definecolor{darkgreen}{rgb}{0.0,0.6,0.0}
\definecolor{purple}{rgb}{.4,.0,.4}
\definecolor{lightgrey}{rgb}{0.7, 0.7, 0.7}
\definecolor{grey}{rgb}{0.4, 0.4, 0.4}

\begin{document}
\title{Hawking radiation from nonrotating singularity-free black holes in conformal gravity }
\author{Jun Zhang} \email{zhangj626@mail2.sysu.edu.cn}\affiliation{School of Physics and Astronomy, Sun Yat-Sen University(Zhuhai Campus), Zhuhai 519082, China}
\author{Yuan Sun} \email{sunyuan@jlu.edu.cn}\affiliation{Center for Theoretical Physics and College of Physics, Jilin University,
Changchun 130012, People’s Republic of China}

\begin{abstract}
We study the sparsity of Hawking radiation from nonrotating singularity-free black holes in conformal gravity. We give a
rigorous bound on the greybody factor for massless scalar field and calculate the sparsity of Hawking radiation from the black
hole. Besides, we investigate the dependence of the bound of the greybody factor and the sparsity of Hawking radiation on the conformal
parameters respectively. Our study shows that when the conformal parameters are large, the increase of conformal parameters will lead to
a more sparse Hawking radiation, while to a less sparse Hawking radiation if the conformal parameters are small.
\end{abstract}
\maketitle

\tableofcontents
\section{Introduction}
Einstein's general relativity is the most elegant and successful gravitational theory subjected to current experimental tests. However, this theory is still imperfect. One of the most important problems of Einstein's gravity is that singularity is inevitable in almost all of its solutions. Since the existence of the singularity in spacetime, causality will break down, and the predicability of the theory will be lost at this point thus to find the solution for the singularity problem becomes theoretically essential. There are many proposals devoted to solving this problem, one attempt is to construct modified gravitational theories which have the properties of conformal invariance~\cite{Englert:1976ep,narlikar:1977nf,Fradkin:1978yw,
Kaku:1982xu,tHooft:2009wdx,Mannheim:2011ds,Bars:2013yba,Prester:2013fia}.
A gravity theory is conformal invariant if its action is invariant under following transformation of metric
\begin{equation} \label{metrictransformation}
{g_{\mu \nu }} \to {g'_{\mu \nu }} = S \left( x \right){g_{\mu \nu }},
\end{equation}
where $S \left( x \right)$ is a regular conformal factor depending on spacetime coordinate. Obviously, Einstein's gravity is not conformal invariant. There are several ways to construct a conformal invariant gravity theory, for example, by introducing an auxiliary field~\cite{Dirac:1973gk} or the conformal invariant Weyl curvature tensor in the action~\cite{Mannheim:2011ds,Mannheim:2016lnx}.

Einstein's gravity is invariant under coordinate transformation, and the singularity at the event horizon can be removed by coordinate transformation therefore this singularity is not intrinsic as to Einstein's gravity. Similarly in the context of conformal gravity theory, since the singularity at $r=0$ can in principle be removed by suitable choice of conformal transformation, hence conformal gravity is free from the singularity.
However, the universe we are living in is not conformal invariant, and if the conformal gravity theory can describe the real universe, then there must be a phase transition at which the conformal symmetry is spontaneously broken. In the broken phase, there are singular and regular metrics with different physical properties due to that the conformal symmetry no longer exists, and it was assumed that our nature has a kind of unknown mechanism to select among these different conformal invariant spacetimes~\cite{Toshmatov:2017kmw}. We believe conformal gravity theory is viable and actually, a class of non-singular black hole spacetime solutions from conformal gravity are compatible with the data from the X-ray observation \cite{Bambi:2017yoz,Zhou:2018bxk}.

In this paper, we will study the nonrotating singularity-free black hole which was found in~\cite{Modesto:2016max,Bambi:2016wdn}. The formation and evaporation process of this black hole has been analyzed in~\cite{Bambi:2016yne}, the quasinormal modes of various kinds of perturbations such as scalar, electromagnetic, axial, and gravitational perturbations in this black hole have been studied in~\cite{Toshmatov:2017bpx,Chen:2019iuo,Liu:2020ddo}. Besides, the violation of energy condition for this black hole has been investigated in~\cite{Toshmatov:2017kmw}.
However, the greybody factor of this black hole is barely explored. So far only the greybody factor of the massless scalar field from this black hole has been numerically studied in \cite{Lin:2020rvv} recently. In this paper, we give a rigorous lower bound for the greybody factor of the massless scalar field radiated from the black hole and study the sparsity of the Hawking radiation. Additionally, we investigate the effects of the variation of conformal factor parameters $N$ and $L$ (see eq.(\ref{S(r)})) on the greybody factor and the sparsity of the Hawking radiation.

The paper is organized as follows. In section.\ref{Review}, we briefly review the nonrotating singularity-free black hole solution obtained in~\cite{Modesto:2016max,Bambi:2016wdn} and the Regge-Wheeler equation for massless scalar field from this black hole derived in~\cite{Toshmatov:2017bpx}. In section.\ref{HG}, we study the power spectrum of Hawking radiation and calculate the rigorous lower bound of the greybody factor for the massless scalar field. Moreover, we give the values for conformal parameters which maximize the lower bound on the greybody factor numerically. Various numerical results for the upper bound of dimensionless parameter $\eta$ used to measure the sparsity of Hawking radiation are given in section.\ref{Sparsity}. Finally, in section.\ref{Summary} we summarize the results of this paper and give a discussion. The geometric units, $\hbar {\rm{ = }}c = {k_{\rm{B}}} = G = 1$, are adopted in this paper.

\section{Non-singular black holes in conformal gravity}\label{Review}
In this section, we briefly review the nonrotating singularity-free black hole solution in conformal gravity~\cite{Modesto:2016max,Bambi:2016wdn}.

The construction of the black hole solution is straightforward and is obtained by simply rescaling the Schwarzschild metric with a conformal factor as follows
\begin{equation} \label{metric}
d{s^2} = S(r)ds_{Schw}^2,
\end{equation}
where $ds_{Schw}^2$ is the line element of Schwarzschild black hole
\begin{equation} \label{Schwarzschild black hole}
ds_{Schw}^2 =  - \left(1-\frac{r_{\text{H}}}{r}\right)d{t^2} + \frac{dr^2}{\left(1-\frac{r_{\text{H}}}{r}\right)} + {r^2}d{\Omega_2 ^2},
\end{equation}
in which $r_{\text{H}}$ is the horizon radius and $S(r)$ is the conformal factor
\begin{equation} \label{S(r)}
S(r) = {\left( {1 + \frac{{{L^2}}}{{{r^2}}}} \right)^{2N}},
\end{equation}
and $N$ is an arbitrary positive integer and $L$ is a new length scale\footnote{Evidently when $N=0$ or $L=0$, the metric (\ref{metric}) reduces to Schwarzschild metric.} which can be either of the order of Planck scale or of the order of the mass $M$ of the black hole. Due to the metric is conformal to the Schwarzschild black hole, the Hawking temperature of the black hole is still $T={1}/{(4\pi r_\text{H})}$. This metric can be the solution of a family of conformally invariant theories, since usually solutions of Einstein's gravity theory are a subset of the wider class solutions of conformally invariant theories \cite{Grumiller:2013mxa,Sultana:2012qp,Said:2014lua}, and indeed, metric in eq.(\ref{metric}) is obtained by the conformal transformation of Schwarzschild metric. It can be verified that the square of the Riemann tensor ${R_{\alpha \beta \gamma \delta }}{R^{\alpha \beta \gamma \delta }}$ of this spacetime is regular everywhere including $r=0$. Besides the spacetime is geodesically complete in the sense that it will take infinite proper time for a particle to reach $r=0$ (for more details see Ref.\cite{Bambi:2016wdn}).

The Klein-Gordon equation for a massless scalar field is
\begin{equation} \label{KG eq}
\frac{1}{{\sqrt { - g} }}{\partial _\mu }({g^{\mu \nu }}\sqrt { - g} {\partial _\nu }\Phi ) = 0.
\end{equation}
Decomposing $\Phi {\rm{ = }}{e^{ - i\omega t}}\frac{{\psi \left( r \right)}}{{r\sqrt {S(r)} }}Y_m^\ell (\theta ,\phi )$ and using the tortoise coordinate $d{r_*} = \frac{{dr}}{{f(r)}}$, the resulting Regge-Wheeler equation is~\cite{Toshmatov:2017bpx}
\begin{equation} \label{Master equation}
\left( {\frac{{{d^2}}}{{d{r_*}^2}} + {\omega ^2} - {V_{{\rm{eff}}}}} \right)\psi \left( r \right) = 0,
\end{equation}
where the effective potential ${{V_{{\rm{eff}}}}}$ is defined as
\begin{equation} \label{effective potential}
{V_{{\rm{eff}}}} \equiv f(r)\left[ {\frac{{\ell \left( {\ell  + 1} \right)}}{{{r^2}}} + \frac{{\frac{d}{{dr}}\left( {f(r)\frac{d}{{dr}}\left( {r\sqrt {S(r)} } \right)} \right)}}{{r\sqrt {S(r)} }}} \right].
\end{equation}

\section{Hawking radiation and Greybody factor}\label{HG}
As discussed in \cite{Gray:2015pma,Miao:2017jtr,Chowdhury:2020bdi}, the energy emitted per unit time for a 4-dimensional black hole is
\begin{equation} \label{energy emit}
\frac{{dE\left( \omega  \right)}}{{dt}} = \sum\limits_\ell  {{\Gamma _\ell }\left( \omega  \right)} \frac{\omega }{{{e^{{\omega  \mathord{\left/
 {\vphantom {\omega  T}} \right.
 \kern-\nulldelimiterspace} T}}} - 1}}\hat{\vec k} \cdot \hat{\vec n}\frac{{{d^3}kdA}}{{{{\left( {2\pi } \right)}^3}}},
\end{equation}
where $dA$ is the infinitesimal surface area, $\hat{\vec k}$ is the unit tangent vector of the momentum of the scalar field, $\hat{\vec{n}}$ is the unit normal vector of $dA$, and ${{\Gamma _\ell }\left( \omega  \right)}$ is the greybody factor. Since the scalar field we consider is massless thus $| {\vec k} | = \omega$, then the energy emitted per unite time on finite surface $A$ is

\begin{equation} \label{energy emit2}
P = \sum\limits_\ell  {\int {{P_\ell }\left( \omega  \right)d\omega } }
\end{equation}
with
\begin{equation} \label{energy emit3}
{P_\ell }\left( \omega  \right) = \frac{A}{{8{\pi ^2}}}\frac{{{\Gamma _\ell }\left( \omega  \right){\omega ^3}}}{{{e^{{\omega  \mathord{\left/
 {\vphantom {\omega  T}} \right.
 \kern-\nulldelimiterspace} T}}} - 1}}.
\end{equation}
It was shown that $A$ can be taken as 27/4 times the horizon area for Schwarzschild black hole to match high frequency results \cite{Page:1976df,Gray:2015pma}, and here we can also take this value, since the cross section in geometrical optical limit for massless particle in this model is the same as that of Schwarzschild black hole~\cite{Lin:2020rvv}. Besides, the Hawking temperature $T$ is independent of conformal parameters $N$ and $L$ (see section.\ref{Review}). Therefore, the dependence of power spectrum ${P_\ell }\left( \omega  \right)$ for Hawking radiation on conformal parameters $N$ and $L$ is as same as that of greybody factor ${\Gamma _\ell }\left( \omega  \right)$.

Calculating the analytical expression for greybody factor is generally impractical and usually, it is obtained from numerical computations~\cite{Harris:2004mf,Harris:2005jx,Kanti:2005ja,Duffy:2005ns}. The analytical calculations are available only in some special cases, such as the low frequency limit~\cite{Unruh:1976fm,Rocha:2009xy,Harmark:2007jy,Kanti:2014dxa,
Anderson:2015bza,Sporea:2015hla}, the near extremal limit~\cite{Cvetic:2009jn,Bredberg:2009pv,Chen:2010bsa,Chen:2010as,Chen:2010ywa,Chen:2010yu}, or the large spacetime dimension limit~\cite{Emparan:2013xia,Emparan:2013moa,Guo:2015swu,Emparan:2020inr}. In this paper, we follow \cite{Visser:1998ke,Boonserm:2008zg,Boonserm:2014rma} to give a rigorous lower bound for the greybody factor. The bound for greybody factor can be written as follow
\begin{equation} \label{Bound 0}
{{\Gamma _\ell }\left( \omega  \right)} \geqslant {\rm{sec}}{{\rm{h}}^2}\left\{ {\int_{ - \infty }^\infty  {\vartheta d{r_*}} } \right\}
\end{equation}
with
\begin{equation} \label{vartheta1 }
\vartheta  = \frac{{\sqrt {{{\left( {h'\left( r \right)} \right)}^2} + {{\left( {{\omega ^2} - {V_{{\rm{eff}}}} - h{{\left( r \right)}^2}} \right)}^2}} }}{{2h\left( r \right)}},
\end{equation}
where $h(r)$ is an arbitrary positive definite function and satisfy the limits $h( - \infty ) = h( + \infty ) = \omega $. A particularly simple
choice of $h(r)$ is $h(r) = \omega$,
which leads to
\begin{equation} \label{Integrate of vartheta 1 }
\int_{ - \infty }^\infty  {\vartheta d{r_*}}  = \int_{2M}^\infty  {\frac{{{V_{{\rm{eff}}}}}}{{2\omega f(r)}}dr}.
\end{equation}
Combining eq.(\ref{effective potential}) and eq.(\ref{Integrate of vartheta 1 }) we have
\begin{equation} \label{Integrate of vartheta 2}
\int_{ - \infty }^\infty  {\vartheta d{r_*}}  = \frac{{2NM(2N - 1)}}{{\omega {L^2}}}\ln \left( {1 + \frac{{{L^2}}}{{4{M^2}}}} \right) + \frac{{N\left( {3N - 2} \right)}}{{{\rm{2}}\omega L}}\left[ {2\arctan \left( {\frac{{2M}}{L}} \right) - \pi } \right] + \frac{{{{\left( {2N - 1} \right)}^2}}}{{8\omega M}} + \frac{{\ell \left( {\ell  + 1} \right)}}{{4\omega M}},
\end{equation}
where $M$ is the mass of the black hole. Substituting the above equation into eq.(\ref{Bound 0}), we obtain the bound as
\begin{equation} \label{Bound 1}
{{\Gamma _\ell }\left( \omega  \right)} \geqslant{\rm{sec}}{{\rm{h}}^2}\left\{ {\frac{{2NM(2N - 1)}}{{\omega {L^2}}}\ln \left( {1 + \frac{{{L^2}}}{{4{M^2}}}} \right) + \frac{{N\left( {3N - 2} \right)}}{{{\rm{2}}\omega L}}\left[ {2\arctan \left( {\frac{{2M}}{L}} \right) - \pi } \right] + \frac{{{{\left( {2N - 1} \right)}^2}}}{{8\omega M}} + \frac{{\ell \left( {\ell  + 1} \right)}}{{4\omega M}}} \right\}.
\end{equation}
Now let us analyze the dependence of this bound on the parameters $N$ and $L$ in various situations. At first,
for $N=0$ we have
\begin{equation} \label{Schwarzschild bound}
{{\Gamma _\ell }\left( \omega  \right)}\geqslant{\rm{sec}}{{\rm{h}}^2}\left\{ {\frac{{\ell \left( {\ell  + 1} \right)}}{{4\omega M}} + \frac{1}{{8\omega M}}} \right\},
\end{equation}
which is exactly the result for Schwarzschild black hole as it should be. Note that this result has been obtained in \cite{Boonserm:2008zg}.
In addition, when $L$ is large, the bound in eq.(\ref{Bound 1}) can be simplified as follows
\begin{equation} \label{Schwarzschild bound1}
{{\Gamma _\ell }\left( \omega  \right)}\geqslant{\text{sec}}{{\text{h}}^2}\left\{ {\frac{{\ell \left( {\ell  + 1} \right)}}{{4\omega M}} + \frac{{{{\left( {2N - 1} \right)}^2}}}{{8\omega M}}} \right\},
\end{equation}
while in small $L$ limit eq.(\ref{Bound 1}) reduces to eq.(\ref{Schwarzschild bound}).

Furthermore, it is not hard to find out specific values for the conformal parameters $N$ and $L$ that can maximize the lower bound obtained in eq.(\ref{Bound 1}). Specifically it can be done as follows. We first rewrite eq.(\ref{Bound 1}) as
\begin{equation} \label{Bound 2}
{\Gamma _\ell }\left( \omega  \right)\geqslant{\text{sec}}{{\text{h}}^2}\left\{ {{{J\left( {L,N,\ell } \right)} \mathord{\left/
 {\vphantom {{h\left( {L,N,\ell } \right)} {\left( {\omega {r_{\text{H}}}} \right)}}} \right.
 \kern-\nulldelimiterspace} {\left( {\omega {r_{\text{H}}}} \right)}}} \right\},
\end{equation}
where we have defined
\begin{equation} \label{h function}
J\left( {L,N,\ell } \right) \equiv \frac{{N(2N - 1)}}{{{{\left( {L/{r_{\text{H}}}} \right)}^2}}}\ln \left( {1 + \frac{{{L^2}}}{{r_{\text{H}}^{\text{2}}}}} \right) + \frac{{N\left( {3N - 2} \right)}}{{{\text{2}}\left( {L/{r_{\text{H}}}} \right)}}\left[ {2\arctan \left( {\frac{{{r_{\text{H}}}}}{L}} \right) - \pi } \right] + \frac{{{{\left( {2N - 1} \right)}^2}}}{4} + \frac{{\ell \left( {\ell  + 1} \right)}}{2}.
\end{equation}
Note the function $y(x) = {\operatorname{sech} ^2}(x)$ is even, monotonously decreasing when $x>0$, and reaches its maximum at $x=0$, thus at fixed $\omega$, the values of $N$ and $L$ that minimize  $\left| {J\left( {L,N,\ell } \right)} \right|$ will maximize ${\text{sec}}{{\text{h}}^2}\left\{ {{{J\left( {L,N,\ell } \right)} \mathord{\left/
 {\vphantom {{h\left( {L,N,\ell } \right)} {\left( {\omega {r_{\text{H}}}} \right)}}} \right.
 \kern-\nulldelimiterspace} {\left( {\omega {r_{\text{H}}}} \right)}}} \right\}$. Therefore to obtain the values for $N$ and $L$ that can maximize the greybody factor we only need to find out the values that can minimize $\left| {J\left( {L,N,\ell } \right)} \right|$.

In TABLE.\ref{tab:table1} and TABLE.\ref{tab:table2}, for given values of ${L \mathord{\left/
 {\vphantom {L {{r_{\text{H}}}}}} \right.
 \kern-\nulldelimiterspace} {{r_{\text{H}}}}}$($N$), we find out numerically the corresponding $N$(${L \mathord{\left/
 {\vphantom {L {{r_{\text{H}}}}}} \right.
 \kern-\nulldelimiterspace} {{r_{\text{H}}}}}$) which can maximize the lower bound obtained in eq.(\ref{Bound 1}).
\begin{table}[hbp]
  \begin{center}
 \caption{For different values of $L/r_{\text{H}}$ the corresponding values of $N$ which maximize the lower bound on greybody factor with $\ell=0$.}
    \label{tab:table1}
    \begin{tabular}{c|c|c|c|c|c|c|c|c|c}
     \hline \hline
     ${L \mathord{\left/
 {\vphantom {L {{r_{\text{H}}}}}} \right.
 \kern-\nulldelimiterspace} {{r_{\text{H}}}}}$&$0.1$&$0.3$&$0.5$&$0.7$&$0.9$&$1.5$&$2$&$2.5$&$3$\\
      \hline
      $N$&8&8&6&3&2&1&1&1&1\\
      \hline\hline
    \end{tabular}
  \end{center}
\end{table}

\begin{table}[h!]
  \begin{center}
 \caption{For different values of $N$ the corresponding values of $L/{r_{\text{H}}}$ which maximize the lower bound on greybody factor with $\ell=0$.}
    \label{tab:table2}
    \begin{tabular}{c|c|c|c|c|c|c|c|c}
     \hline \hline
     $N$&$1$&$3$&$5$&$7$&$9$&$12$&$15$&$20$\\
      \hline
      ${L \mathord{\left/
 {\vphantom {L {{r_{\text{H}}}}}} \right.
 \kern-\nulldelimiterspace} {{r_{\text{H}}}}}$&2.893&0.785&0.558&0.456&0.395&0.337&0.299&0.257\\
 \hline\hline
    \end{tabular}
  \end{center}
\end{table}

\begin{figure}
\centering
\subfigure{
\includegraphics[width=5cm]{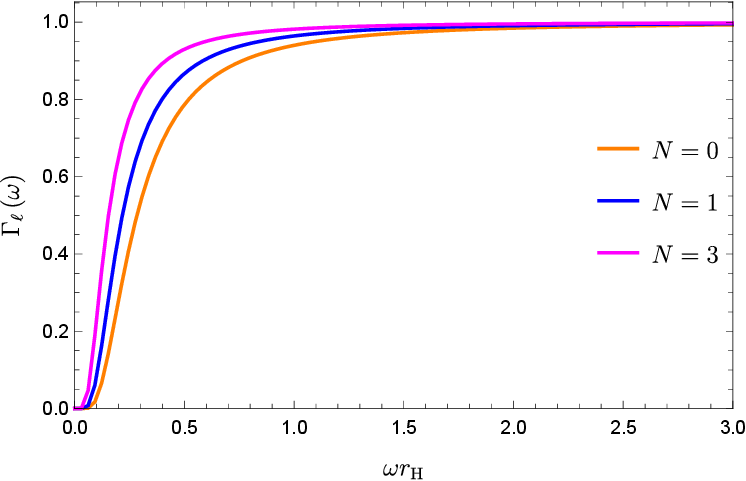}
}
\subfigure{
\includegraphics[width=5cm]{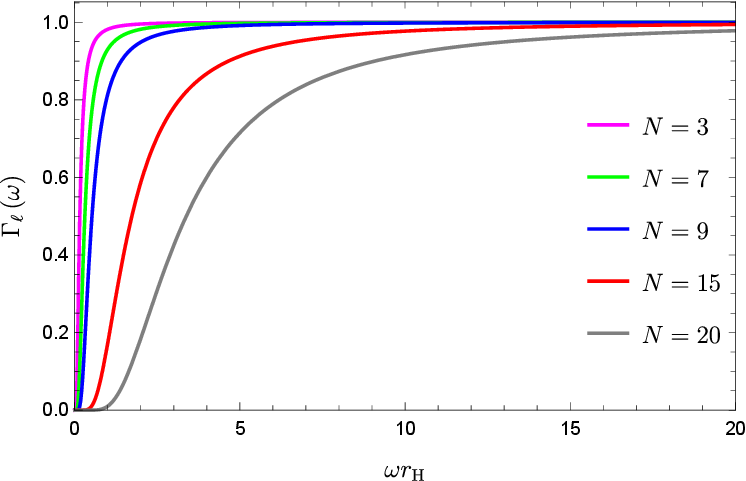}
}
\\
\subfigure{
\includegraphics[width=5cm]{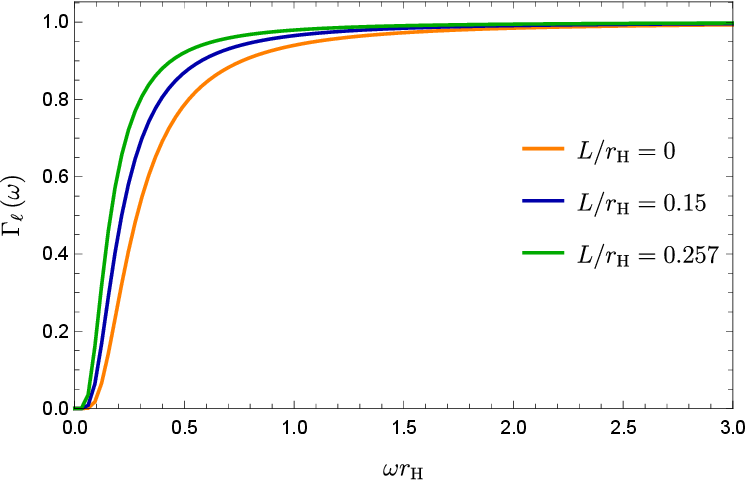}
}
\subfigure{
\includegraphics[width=5cm]{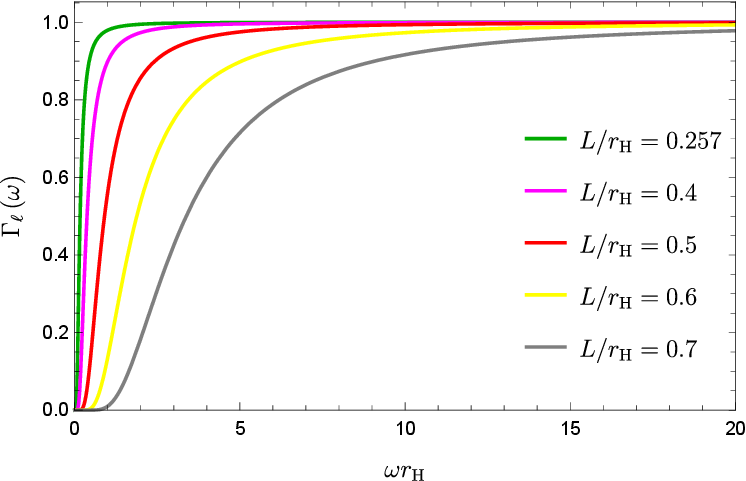}
}
\DeclareGraphicsExtensions.
\caption{(Top) Plots of the lower bound of Greybody factor ${{\Gamma _\ell }\left( \omega  \right)}$ as a function of $\omega r_\text{H}$ for different values of $N$ with $\ell=0$ and ${L}/{r_\text{H}}=0.7$; (Bottom) Plots of the bound of Greybody factor ${{\Gamma _\ell }\left( \omega  \right)}$ as a function of $\omega r_\text{H}$ for different values of ${L}/{r_\text{H}}$ with {$\ell=0$} and $N=20$.}
\label{figure.1}
\end{figure}

To see the dependence of greybody factor and power spectrum of Hawking radiation on conformal parameters $N$, $L$, and the frequency $\omega$, intuitively, we plot the lower bound of greybody factor and power spectrum of Hawking radiation as a function of frequency for different values of $N$ and $L$. FIG.\ref{figure.1} shows that, when the value of ${L \mathord{\left/{\vphantom {L {{r_{\rm{H}}}}}} \right.\kern-\nulldelimiterspace} {{r_{\rm{H}}}}}$ is fixed, the increase of $N$ leads the lower bound of the greybody factor to increase at first (top left) and then to decrease (top right). Similarly the increase of  ${L \mathord{\left/{\vphantom {L {{r_{\rm{H}}}}}} \right.\kern-\nulldelimiterspace} {{r_{\rm{H}}}}}$ will enhance the lower bound of the greybody factor at the beginning (bottom left) and suppress it later (bottom right) for a given value of $N$. Obviously FIG.\ref{figure.1} is consistent with TABLE.\ref{tab:table1} and TABLE.\ref{tab:table2}.

\begin{figure}
\centering
\subfigure{
\includegraphics[width=5.5cm]{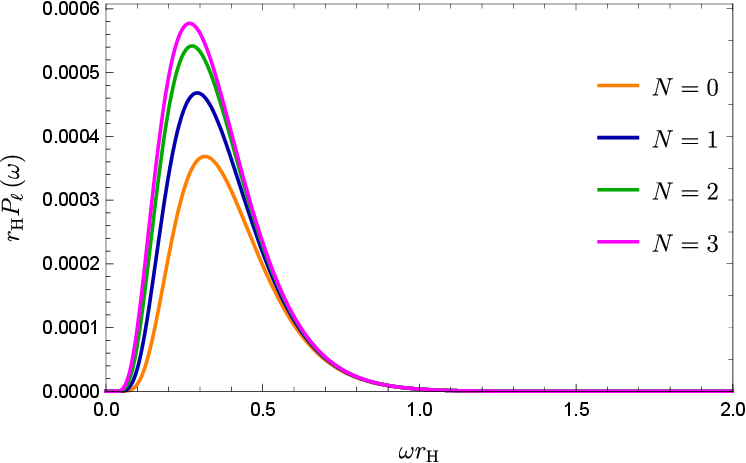}
}\subfigure{
\includegraphics[width=5.5cm]{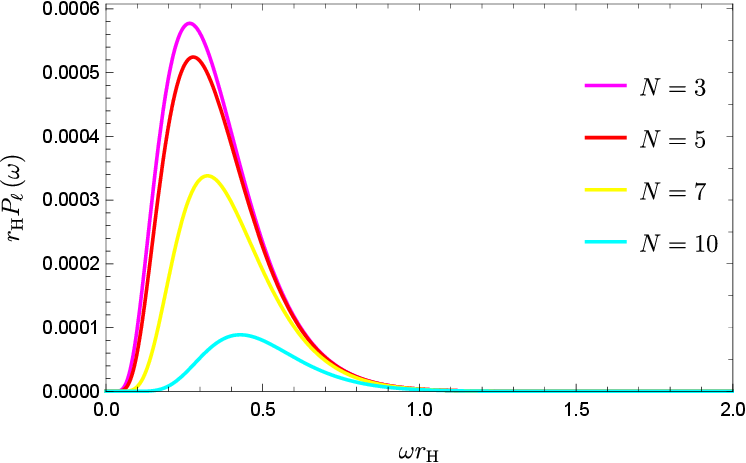}
}
\subfigure{
\includegraphics[width=5.5cm]{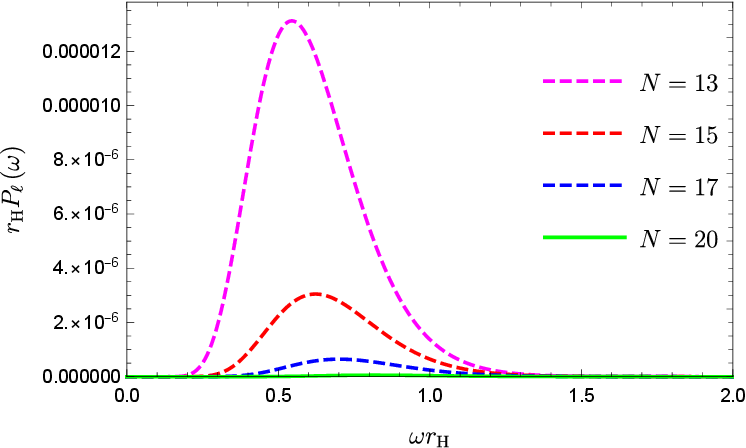}
}
\\
\subfigure{
\includegraphics[width=5.5cm]{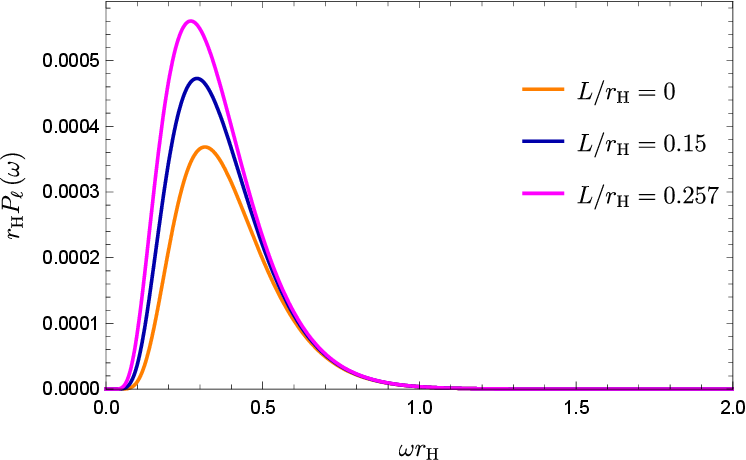}
}
\subfigure{
\includegraphics[width=5.5cm]{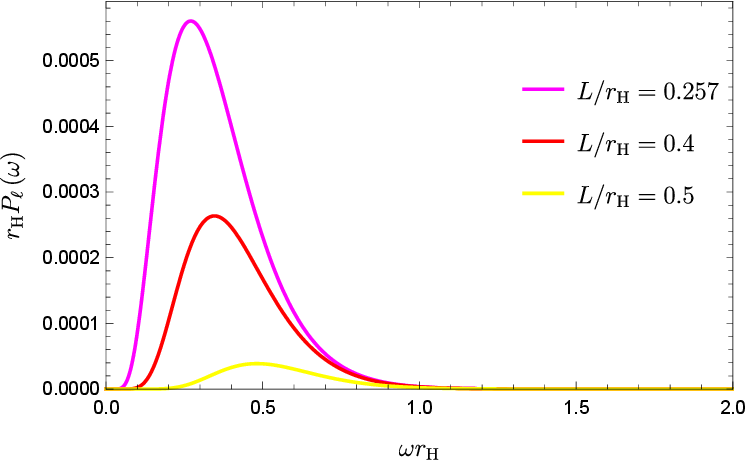}
}
\subfigure{
\includegraphics[width=5.5cm]{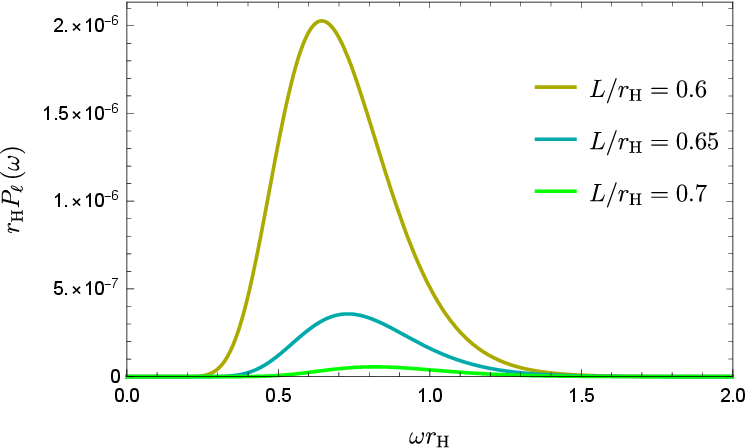}
}
\DeclareGraphicsExtensions.
\caption{(Top) Plots of lower bounds on the power spectrum ${r_{\rm{H}}}{P_\ell }\left( \omega  \right)$ as a function of $\omega r_\text{H}$ for different values of $N$ with $\ell=0$ and ${L}/{r_\text{H}}=0.7$; (Bottom) Plots of lower bounds on the power spectrum} ${r_{\rm{H}}}{P_\ell }\left( \omega  \right)$ as a function of $\omega r_\text{H}$ for different values of ${L}/{r_\text{H}}$ with {$\ell=0$} and $N=20$.
\label{figure.2}
\end{figure}

Furthermore, as shown in FIG.\ref{figure.2}, for a fixed value of $L/{r_{\rm H}}$, the peak value of lower bound on the power spectrum of Hawking radiation increases as the increase of $N$ in the first place and decreases if $N$ continue to grow. While from the bottom of FIG.\ref{figure.2} one can find that for a relatively small value of $L/{r_{\rm H}}$ the growth of  $L/{r_{\rm H}}$ will cause the peak value of the lower bound on the Hawking radiation power spectrum to increase. Once $L/{r_{\rm H}}$ becomes relatively larger the increase of $L/{r_{\rm H}}$ will give rise to the decrease of the peak value of the lower bound of the power spectrum when $N$ is fixed. Also, we find that the values of $L/{r_{\rm H}}$ and $N$ which maximize the lower bound of the power spectrum of Hawking radiation are exactly those values for $L/{r_{\rm H}}$ and $N$ that maximize the lower bound of the greybody factor, which is in agreement with our analysis about the dependence of the lower bound on the Hawking radiation's power spectrum on $N$ and $L$ above.


\begin{figure}
\centering
\subfigure{
\includegraphics[width=5.5cm]{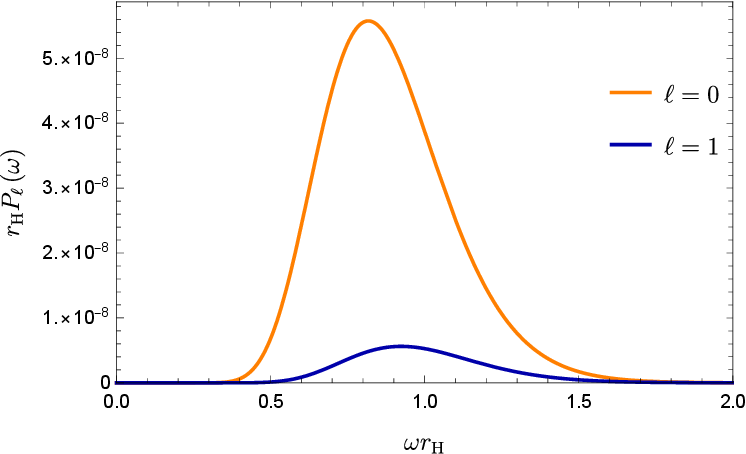}
}
\subfigure{
\includegraphics[width=5.5cm]{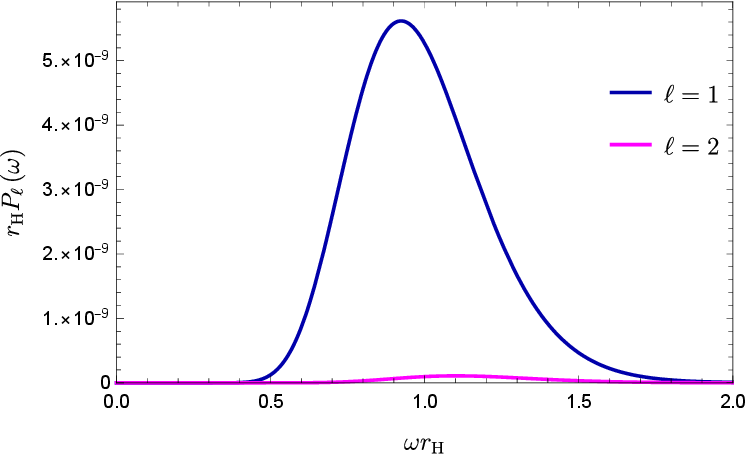}
}
\subfigure{
\includegraphics[width=5.5cm]{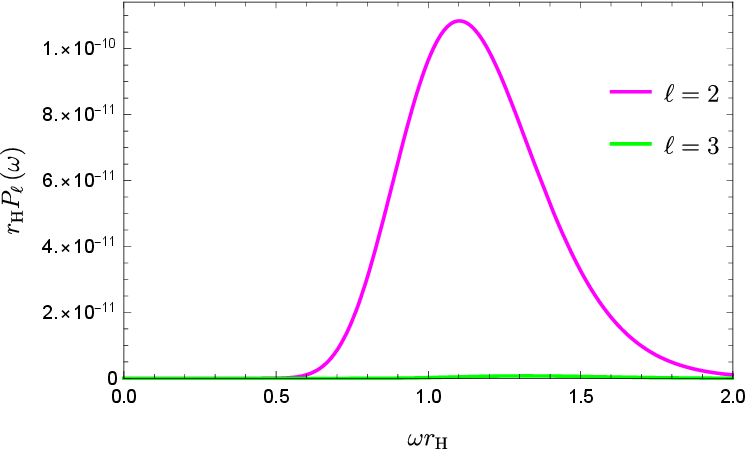}
}
\DeclareGraphicsExtensions.
\caption{Plots of lower bounds on the power spectrum ${r_{\rm{H}}}{P_\ell }\left( \omega  \right)$ as a function of $\omega r_\text{H}$ for different values of $\ell$ with $N=20$ and ${L}/{r_\text{H}}=0.7$.}
\label{figure.3}
\end{figure}

\section{The sparsity of Hawking radiation}\label{Sparsity}
The sparsity of Hawking radiation can be measured by the following quantity~\cite{Gray:2015pma}
\begin{equation} \label{Sparsity parameter}
\eta  = \frac{{{\tau _{{\rm{gap}}}}}}{{{\tau _{{\rm{emission}}}}}},
\end{equation}
where ${{\tau _{{\rm{gap}}}}}$ is the average time interval between the emission of two Hawking quanta, which is proportional to inverse of the total Hawking radiation power $P$ in eq.(\ref{energy emit2}). More specifically, it can be computed by~\cite{Gray:2015pma,Paul:2016xvb}
\begin{equation} \label{Average time}
{\tau _{{\rm{gap}}}} = \frac{{{\omega _{{\rm{peak}}}}}}{P},
\end{equation}
here ${{\omega _{{\rm{peak}}}}}$ is the value at which the power spectrum $\sum\limits_\ell  {{P_\ell }\left( \omega  \right)} $ in eq.(\ref{energy emit2}) is maximized.  Before computing $\omega_{\text{peak}}$, let's show a fact that in eq.(\ref{energy emit2}), we can make an approximation that we only need to consider the term with $\ell=0$ in the summation.\footnote{We thank the anonymous referee for pointing out this point. } The justification of this approximation can be obtained by the observation of FIG.\ref{figure.3} . Obviously, we can ignore the Hawking radiation modes with $\ell \neq 0 $.  So we can have
\begin{equation}
P \approx \int {{P_\ell }} \left( \omega  \right)d\omega {\text{  }}\left( {\ell  = 0} \right),
\end{equation}
and naturally,  we can choose the value that maximizes ${P_\ell }\left( \omega  \right){\text{  }}\left( {\ell  = 0} \right)$ (In what follows, we will simply express it as ${{P_0}\left( \omega  \right)}$.) for $\omega_{\text{peak}}$.
 For different parameters $N$ and $L$, the values we found for ${{\omega _{{\text{peak}}}}}$ are shown in the TABLE.\ref{tab:table5} and TABLE.\ref{tab:table6}.\footnote{In fact, the values we found for ${{\omega _{{\rm{peak}}}}}$ in the TABLE.\ref{tab:table5} and TABLE.\ref{tab:table6} maximize the lower bound of the power spectrum ${{P_0}\left( \omega  \right)}$ instead of ${{P_0}\left( \omega  \right)}$ itself. The argument for doing this is as follows.
 We attempt to study the effect of $N$ and $L$ on sparsity, and to realize this goal, eq.(\ref{etamax}) tells us that we need to analyze the corresponding effects on $\omega_{\text{peak}}$ and ${\int {{P_0}\left( \omega  \right)d\omega } }$, respectively. From eq.(\ref{energy emit3}), we know that the $N$ and $L$ dependence of ${{P_0}\left( \omega  \right)}$ only comes from the greybody factor. On the other hand, $\omega_{\text{peak}}$ is the frequency that maximizes ${{P_0}\left( \omega  \right)}$, so its dependence on $N$ and $L$ is also from the greybody factor. Therefore, as long as we can obtain the dependence of the greybody factor on $N$ and $L$, we can then find the relationship between sparsity and parameters $N$ and $L$.   However, the exact value of greybody factor is inaccessible so far. To make progress with difficulty, we choose to do a qualitative study by assuming that the qualitative effect of $N$ and $L$ on the  greybody factor is the same as that on its bound i.e.when the change of $N$ and $L$ leads to the greybody factor increase or decrease, we expect that its bound would also increase or decrease accordingly. By using this assumption, we then can have a chance to know the qualitative behavior of the sparsity under the variation of $N$ and $L$ by the argument above. Notice that a similar assumption has also been made implicitly in\cite{Miao:2017jtr}\cite{Chowdhury:2020bdi}. Of course, further justification is needed to support this assumption, and we hope to study this problem in the future.}

\begin{table}[h!]
  \begin{center}
 \caption{Numerical values of  ${\omega _{{\text{peak}}}}{r_H}$ for different values of $N$  with {$\ell=0$} and ${L}/{{{r_{\rm{H}}}}}=0.7$.}
    \label{tab:table5}
    \begin{tabular}{c|c|c|c|c|c|c|c|c|c|c|c|c|c}
     \hline \hline
     $N$&$0$&$1$&$2$&$3$&$4$&$5$&$6$&$7$&$8$&$9$&$10$&$15$&$20$\\
      \hline
      ${\omega _{{\text{peak}}}}{r_H}$&0.316&0.291&0.275&0.267&0.268&0.279&0.297&0.324&0.356&0.391&0.428&0.623&0.817\\
      \hline\hline
    \end{tabular}
  \end{center}
\end{table}

\begin{table}[h!]
  \begin{center}
 \caption{Numerical values of ${\omega _{{\text{peak}}}}{r_H}$ for different values of ${L}/{{{r_{\rm{H}}}}}$ with {$\ell=0$} and $N=20$.}
    \label{tab:table6}
    \begin{tabular}{c|c|c|c|c|c|c|c|c|c|c|c}
     \hline \hline
     ${L \mathord{\left/
 {\vphantom {L {{r_{\rm{H}}}}}} \right.
 \kern-\nulldelimiterspace} {{r_{\rm{H}}}}}$&0&0.05&0.1&0.15&0.2&0.257&0.3&0.35&0.4&0.5&0.7\\
      \hline
      ${\omega _{{\text{peak}}}}{r_H}$&0.316&0.312&0.303&0.290&0.278&0.271&0.276&0.301&0.346&0.482&0.817\\
      \hline\hline
    \end{tabular}
  \end{center}
\end{table}

\begin{table}[h!]
  \begin{center}
 \caption{Numerical values of dimensionless parameter ${\eta _{\max }}$ for different values of $N$  with {$\ell=0$} and ${L}/{{{r_{\rm{H}}}}}=0.7$.}
    \label{tab:table3}
    \begin{tabular}{c|c|c|c|c|c|c|c|c|c|c|c|c|c}
     \hline \hline
     $N$&$0$&$1$&$2$&$3$&$4$&$5$&$6$&$7$&$8$&$9$&$10$&$15$&$20$\\
      \hline
      ${\eta _{\max }}$&122.8&83.0&63.9&56.6&57.7&67.8&$91.6$&$140.3$&$238.5$&$440.6$&$868.2$&$45850$&$3.85\times10^{6}$\\
      \hline\hline
    \end{tabular}
  \end{center}
\end{table}
\begin{table}[h!]
  \begin{center}
 \caption{Numerical values of dimensionless parameter ${\eta _{\max }}$ for different values of ${L}/{{{r_{\rm{H}}}}}$ with {$\ell=0$} and $N=20$.}
    \label{tab:table4}
    \begin{tabular}{c|c|c|c|c|c|c|c|c|c|c|c}
     \hline \hline
     ${L \mathord{\left/
 {\vphantom {L {{r_{\rm{H}}}}}} \right.
 \kern-\nulldelimiterspace} {{r_{\rm{H}}}}}$&0&0.05&0.1&0.15&0.2&0.257&0.3&0.35&0.4&0.5&0.7\\
      \hline
      ${\eta _{\max }}$&122.8&116.4&100.3&81.5&66.7&59.9&65.6&96.4&202.9&2407.02&$3.85\times10^{6}$\\
      \hline\hline
    \end{tabular}
  \end{center}
\end{table}
${{\tau _{{\rm{emission}}}}}$ is the characteristic time for an individual quantum emitted from black hole, which can be taken equal to the period of the quantum emitted, since the quanta emitted at peak value can only be temporarily localized within a few oscillation periods. Therefore $\tau _{{\text{emission}}}$ obeys the bound
\begin{equation} \label{characteristic time}
{\tau _{{\text{emission}}}} \geqslant \frac{{2\pi }}{{{\omega _{{\text{peak}}}}}},
\end{equation}
which leads to an upper bound for sparsity
\begin{equation} \label{etamax}
{\eta  \leqslant \frac{{\omega _{{\text{peak}}}^2}}{{2\pi P}}\equiv {\eta _{\max }} \approx \frac{{\omega _{{\text{peak}}}^2}}{{2\pi \int {{P_0}\left( \omega  \right)d\omega } }}.}
\end{equation}
 Since the exact value for greybody factor is unknown so far, we will use its lower bound in (\ref{Bound 1}) to calculate ${P_0}\left( \omega  \right)$ in (\ref{etamax}), and we have given our corresponding argument in the footnote (3).

In summary, the sparsity of radiation in eq.(\ref{Sparsity parameter}) tells us that if  $\eta  \gg 1$, i.e. the average interval time $\tau _{{\rm{gap}}}$ between the emission of two quanta is much larger than the time $\tau _{{\text{emission}}}$  taken by an individual quantum emitted from the black hole, then naturally the Hawking radiation would be very sparse. On the contrary, if $\eta  \ll 1$  we will have almost continuous Hawking radiation.

In what follows we will study the dependence of upper bound of sparsity on the conformal parameters $N$ and $L$ numerically.
Firstly, let us investigate the behavior of $\eta_\text{max}$ with different $N$ while keeping $L$ fixed. In TABLE.\ref{tab:table3} the values  of ${\eta _{\max }}$ for different values of $N$ with $L$ fixed are listed.
The TABLE.\ref{tab:table3} shows that given ${L \mathord{\left/
 {\vphantom {L {{r_{\text{H}}}}}} \right.
 \kern-\nulldelimiterspace} {{r_{\text{H}}}}}=0.7$ and $\ell=0$, as $N$ increasing, the values of ${\eta _{\max }}$ will decreases at first, and begin to increase at $N=3$.
In the next step, we go on to study the case with fixed $N=20$ (also $\ell=0$) and letting $L/r_H$ vary. The results are presented in TABLE.\ref{tab:table4}, from which it is clear that the situation in TABLE.\ref{tab:table4} is similar to that of TABLE.\ref{tab:table3}.  TABLE.\ref{tab:table4} shows that ${\eta _{\max }}$ takes its minimum value at ${L \mathord{\left/
 {\vphantom {L {{r_{\text{H}}}}}} \right.
 \kern-\nulldelimiterspace} {{r_{\text{H}}}}}=0.257$, and as ${L \mathord{\left/
 {\vphantom {L {{r_{\text{H}}}}}} \right.
 \kern-\nulldelimiterspace} {{r_{\text{H}}}}}$ getting larger than this values, ${\eta _{\max }}$  will increase as the increase of ${L \mathord{\left/
 {\vphantom {L {{r_{\text{H}}}}}} \right.
 \kern-\nulldelimiterspace} {{r_{\text{H}}}}}$.

In TABLE.\ref{tab:table3} and TABLE.\ref{tab:table4} the sparsity of $N=0$ or ${L \mathord{\left/
 {\vphantom {L {{r_{\text{H}}}}}} \right.
 \kern-\nulldelimiterspace} {{r_{\text{H}}}}}=0$ corresponds to Schwarzschild black hole, we show that when the $N$ and $L$ are relatively small (comparing with corresponding values that minimize the sparsity), one can have nonrotating singularity-free black holes whose Hawking radiation are less sparse than that of Schwarzschild black hole with the same mass.

\section{Summary}\label{Summary}
We studied the sparsity of Hawking radiation from a nonrotating singularity-free black hole in conformal gravity proposed in \cite{Modesto:2016max,Bambi:2016wdn}. We mainly focused on analyzing the dependence of the bound of the sparsity of Hawking radiation on the conformal parameters $N$ and $L$ (see eq.(\ref{S(r)})).

We first obtained a rigorous lower bound on the greybody factor (see eq.(\ref{Bound 1})) for the massless scalar field radiated from the black hole and analyzed the dependence of the greybody factor on the conformal parameters $N$ and $L$ via the bound. We found that there exist values for conformal parameters $N$ and $L$ that can maximize the lower bound on the greybody factor. For fixed $L$ (or $N$), the values of $N$ (or $L$)  maximizing this bound are given numerically (see TABLE.\ref{tab:table1} and TABLE.\ref{tab:table2}). Furthermore, we investigated the power spectrum of Hawking radiation (see eq.(\ref{energy emit3})) and showed that the dependence of power spectrum for Hawking radiation on conformal parameters $N$ and $L$ is the same as that of the greybody factor.

Finally, with the rigorous lower bound on greybody factor and power spectrum for Hawking radiation in hand, we studied the sparsity of Hawking radiation by the upper bound ${\eta _{\max }}$ (see eq.(\ref{etamax})) of dimensionless parameter $\eta$ (see eq.(\ref{Sparsity parameter})). We discovered that when the conformal parameters $N$ and $L$ are relatively small comparing with the values that minimize the sparsity one can have nonrotating singularity-free black holes with Hawking radiation less sparse than that of Schwarzschild black hole with the same mass (see TABLE.\ref{tab:table3} and TABLE.\ref{tab:table4}). When the values of conformal parameters surpass the values that minimize the sparsity of Hawking radiation, the further increase of conformal parameters will enhance ${\eta _{\max }}$ rapidly, which implies the Hawking radiation becomes sparser.

We are unable to identify whether the Hawking radiation from this black hole is sparse or not since in this paper, we only obtained the upper bounds of $\eta$. If one wants to know whether the Hawking radiation from this black hole is sparse or not, one can try to find out the lower bound of $\eta$ defined in eq(\ref{Sparsity parameter}). This is an open question which we would like to address in the future. Besides, we plan to extend this study to the case of rotating singularity-free black hole which has also been proposed in \cite{Modesto:2016max,Bambi:2016wdn}, since rotating black holes are more likely to exist in our universe.

\newpage
\section*{Acknowledgement}
We would like to thank Jia-Rui Sun, Bin Wu, and Zhen-Ming Xu for useful discussions and the anonymous referee for helpful comments and suggestions. This work was supported by the National Natural Science Foundation of China (No.~11675272 and No.~12105113).

\end{document}